\documentclass[preprint,aps,prb]{revtex4-2}
\usepackage[utf8]{inputenc}

\usepackage{xcolor}   %textcolor command
\usepackage{graphicx} %includegraphics command
\usepackage{tikz}     %tikzpicture environment
\usepackage{physics}  %symbol bra, ket, ...
\usepackage{amsfonts} %special fonts

\newcommand{\Co}[1]{a^\dag_{#1}}
\newcommand{\Ao}[1]{a^{\phantom\dag}_{#1}}

\newcommand{\CFo}[1]{\hat{\psi}^\dag\left(\vb{#1}\right)}
\newcommand{\AFo}[1]{\hat{\psi}\left(\vb{#1}\right)}

\newcommand{\DF}[1]{\mathbf{{D}}^{#1}}
\newcommand{\DFMixed}{\mathbf{{D}_{\text{m}}}}

\newcommand{\DFM}[1]{\mathbf{{A}}^{#1}}
\newcommand{\DFMN}[2]{\mathbf{{A}}^{#1}_{#2}}
\newcommand{\DFm}[2]{{A}^{#1}_{#2}}
\newcommand{\DFMM}[3]{\mathbf{\left[A^{#1}\right]}_{#2,#3}}
\newcommand{\DFmm}[3]{\left[A^{#1}_{#2}\right]_{#3}}
\newcommand{\DFMMixed}{\mathbf{{A}^{\text{m}}}}

\newcommand{\DFP}[1]{\mathbf{{C}}^{#1}}
\newcommand{\DFPN}[2]{\mathbf{{C}}^{#1}_{#2}}
\newcommand{\DFp}[2]{{C}^{#1}_{#2}}
\newcommand{\DFPM}[3]{\mathbf{\left[C^{#1}\right]}_{#2,#3}}
\newcommand{\DFpm}[3]{\left[C^{#1}_{#2}\right]_{#3}}
\newcommand{\DFPMixed}{\mathbf{{C}^{\text{m}}}}

\newcommand{\BigSet}{\mathcal{A}_{N}\times\mathcal{C}_{N}}      %The set of big pair of matrices
\newcommand{\BigSetM}{\mathcal{A}_{N-1}\times\mathcal{C}_{N-1}} %The set of big pair of matrices
\newcommand{\BigSetQ}{\mathcal{A}_{N}/U\left[\binom{n}{N-1}\right]\times\mathcal{C}_{N}/U\left[\binom{n}{N+1}\right]}   %The quotient set of big pair of matrices
\newcommand{\BigSetMQ}{\mathcal{A}_{N-1}/U\left[\binom{n}{N-2}\right]\times\mathcal{C}_{N-1}/U\left[\binom{n}{N}\right]} %The quotient set of big pair of matrices
\newcommand{\SetQ}{\mathcal{Q}_N}      %Short name of \BigSetQ
\newcommand{\SetMQ}{\mathcal{GQ}_{N-1}} %Short name of \BigSetMQ

\newcommand{\PF}[1]{\mathbf{P}^{#1}}                 %ACMP in \BigSet
\newcommand{\GPF}[1]{\mathcal{G}^{#1}_{N-1}}         %set of n ACMPs for the 2-RDM
\newcommand{\GPFMixed}{\mathcal{G}^{\text{m}}_{N-1}} %mixed-state of a set of n ACMPs for the 2-RDM

\newcommand{\Hone}[1]{h_{#1}}
\newcommand{\Htwo}{H^{ii'}_{jj'}}
\newcommand{\HTone}[1]{\tilde{h}_{#1}}
\newcommand{\HTtwo}{\tilde{H}^{ii'}_{jj'}}
\newcommand{\ORDM}{$1$-RDM}
\newcommand{\RDMOne}{\mathbf{{}^1\Gamma}}
\newcommand{\RDMone}[1]{{}^1\Gamma_{#1}}
\newcommand{\RDMOneMixed}{\mathbf{{}^1\Gamma^{\text{m}}}}
\newcommand{\TRDM}{$2$-RDM}
\newcommand{\RDMTwo}{\mathbf{{}^2\Gamma}}
\newcommand{\RDMtwo}{{}^2\Gamma^{ii'}_{jj'}}
\newcommand{\RDMTwoMixed}{\mathbf{{}^2\Gamma^{\text{m}}}}
\newcommand{\RDMtwoMixed}{\left[\RDMTwoMixed\right]^{ii'}_{jj'}}

\newcommand{\pscal}[2]{\langle#1,#2\rangle} %Scalar product
\newcommand{\Trans}[1]{#1{}^\dag} %Transposition
\renewcommand{\Trace}[1]{Tr\!\left(#1\right)} %Trace

\newcommand{\alert}[1]{\textcolor{red}{#1}}

\begin{document}

\title{Exact solution of the many-body problem with a $\mathcal{O}\left(n^6\right)$~complexity}
\author{Thierry Deutsch}
\email{Thierry.Deutsch@cea.fr}
\affiliation{University Grenoble Alpes,CEA, IRIG-MEM, Grenoble, France}
\date{\today}

\begin{abstract}
    In this article, we define a new mathematical object, called a pair~$\left(\DFM{},\DFP{}\right)$ of anti-commutation matrices (ACMP) based on the anti-commutation relation~$\Co{i}\Ao{j} + \Ao{j}\Co{i} = \delta_{ij}$ applied to the scalar product between the many-body wavefunctions.
    
    This ACMP~$\left(\DFM{},\DFP{}\right)$ explicitly separates the different levels of correlation.
    The one-body correlations are defined by a ACMP~$\left(\DFM{0},\DFP{0}\right)$ and the two-body ones by a set of $n$ ACMPs~$\left(\DFM{i},\DFP{i}\right)$ where $n$ is the number of states.
    We show that we can have a compact and exact parametrization with $n^4$ parameters of the two-body reduced density matrix (\TRDM) of any pure or mixed $N$-body state to determine the ground state energy with a $\mathcal{O}\left(n^6\right)$~complexity.
\end{abstract}

\maketitle

\section{Introduction}

The Coulson's challenge~\cite{Coulson1960a} has the goal to use only the two-body reduced density matrix~(\TRDM) to minimize the total energy of an electronic system. The \TRDM\ is a compact mathematical object which can only be described with~$n^4$ parameters, where $n$ is the number of states, but many inequality conditions are necessary in order to have a $N$-representable \TRDM~\cite{Coleman.1963} \textit{i.e.} coming from a $N$-body anti-symmetrized wavefunction~$\ket{\Psi_N}$. On contrary, the wavefunction $\ket{\Psi_N}$ is not a compact mathematical object which can be expressed as a sum of~$\binom{n}{N}$ Slater determinants, a number varying exponentially versus the number of electrons~$N$, but has the nice property to form a Hilbert space, easy to manipulate.

John A. Coleman~\cite{Coleman-Yukalov.2000,Coleman.2001} and many others tried to define all conditions to constrain a \TRDM~to be $N$-representable. 
There were also many attempts to define some very accurate approximations based on the one-body density matrix (\ORDM)~\cite{Valdemoro.1992} or on the contracted Schr\"odinger equation~\cite{Mazziotti.1998,Valdemoro2007}. 
The bottleneck is the large number of inequality conditions to check in order to have an exact $N$-representable \TRDM, which were all determined by D.~Mazziotti~\cite{Mazziotti2012} in 2012, based on many-body operators.
The semi-definite programming~\cite{Maho-Nakatsuji-Ehara-2001,Mazziotti.PRL.2004,PhysRevA.72.052505} is in this case used to impose some constraints.
This approach has been explored to calculate the energies of molecules~\cite{Gidofalvi2005} and more recently in solid state physics~\cite{Mazziotti2021}.

The goal of this article is to give a new formalism based on a more compact object than a wavefunction with the main idea that the \TRDM\ is interpreted as transitions between wavefunctions in the Fock space similar to the Dyson orbital approach~\cite{Ortiz2020} or the Feynman-Dyson amplitudes~\cite{Morrison-Ayers1995}.
The structure of the paper is as follows: We start by introducing a new mathematical object~$\DF{}=\left(\DFM{},\DFP{}\right)$, called a pair of anti-commutation matrices (ACMP) which encodes in a new way the anti-commutation relation.
The ACMPs are only defined with equality constraints translating the anti-commutation relations into linear matrix algebra.
These equality constraints are necessary by construction.

In a second part, we show that these conditions are sufficient, any matrix pair which respects these conditions are pure-state or mixed-state $N$-representable.
Finally, because we are only looking for the solutions of a two-body Hamiltonian, we show that we can compact these ACMP keeping the $N$-representability based only on the equality constraints.

Then we develop the Lagrangian applying the equality constraints to have ACMPs by means of Lagrange multipliers.
We give some preliminary numerical examples and conclude by a general discussion.

\section{Building the anti-commutation matrices}

\subsection{The Hamiltonian and the reduced density matrices}

$n$ possible states are associated to the corresponding creation and annihilation operators~$\Co{i}$ and~$\Ao{i}$ and a two-body Hamiltonian~$\hat{H}=\hat{h}_1 + \hat{H}_2$ where $\hat{h}_1$ is the one-body part and $\hat{H}_2$ the two-body one, which is the general case to calculate the electronic structures of any atomic system.
We are looking for the lowest total energy of a $N$ electron system with the corresponding ground state wavefunction~$\ket{\Psi^0_{N}}$, normalized to $1$, given by $E_0 = \mel{\Psi^0_{N}}{\hat{H}}{\Psi^0_{N}}$.
Because the physical interactions are only a sum of two-body interactions, the total energy can be expressed using only the one-body reduced density matrix~$\RDMOne$ (\ORDM) and the two-body reduced density matrix~$\RDMTwo$ (\TRDM) as
\begin{align}
    \hat{h}_1 & = \sum_{ii'}^n \Hone{ii'} \Co{i}\Ao{i'}, \quad \hat{H}_2 = \sum_{ij;i'j'}^n \Htwo \Co{i}\Co{j}\Ao{j'}\Ao{i'}\\
    E & = \sum_{ii'}^n \Hone{ii'} \mel {\Psi^0_N} {\Co{i}\Ao{i'}} {\Psi^0_N} + \sum_{ij;i'j'}^n H^{ii'}_{jj'} \mel {\Psi^0_N} {\Co{i}\Co{j}\Ao{j'}\Ao{i'}} {\Psi^0_N} \\
      & = \sum_{ii'}^n \Hone{ii'} \RDMone{ii'} + \sum_{ij;i'j'}^n H^{ii'}_{jj'} \RDMtwo\\
\text{with}\; & \RDMone{ii'}  = \mel {\Psi^0_N} {\Co{i}\Ao{i'}} {\Psi^0_N} \;\text{and}\;\RDMtwo = \mel {\Psi^0_{N}} {\Co{i}\Co{j}\Ao{j'}\Ao{i'}} {\Psi_{N}^0}.
\end{align}
It is possible to generalize to the field operators of creation~$\CFo{x}$ and annihilation~$\AFo{x}$ but the advantage of starting with a discrete and finite number of states is that the linear algebra can be used to give some clues about the structure of the object we propose.
We prefer to use the notation with the indices~$i$ and~$i'$ in superscript and the two other interior indices~$j$ and~$j'$ in subscript which is more intuitive reflecting the interpretation we propose later in term of the $j$th column of a matrix indexed by~$i$.

The structure of the Mathematics for fermions is governed by the relations of anti-commutation as
\begin{equation}
\Ao{i}\Ao{j} + \Ao{j}\Ao{i} = 0, \quad
\Co{i}\Co{j} + \Co{j}\Co{i} = 0, \quad
\Co{i}\Ao{j} + \Ao{j}\Co{i} = \delta_{ij} \label{rel-ca}.
\end{equation}

The Hilbert space~$\mathcal{H}_N$ of the $N$-body wavefunctions $\ket{\Psi_{N}}$ is a vector space associated with a scalar product by definition.
In particular, there exists a natural basis set of~$\mathcal{H}_N$ which are the Slater determinant~$\ket{I_N}$, anti-symmetric by construction, where $I_N$ is an index of $N$ ordered numbers $\left(i_1,\ldots,i_N\right)$ taken into $n$ numbers which corresponds to the $N$ occupied states of the $N$-body wavefunction 
$\ket{i_1,\ldots,i_N}=\Co{i_N}\ldots\Co{i_1}\ket{0}$ where $\ket{0}$ is the $0$-body wavefunction. The number of Slater determinants scales exponentially as~$\binom{n}{N}$ versus the number~$N$ of electrons.
Any wavefunction $\ket{\Psi_N}$ can be decomposed into all Slater determinants as 
\begin{equation}
    \ket{\Psi_N} = \sum_{I_N} c_{I_N} \ket{I_N}
\end{equation}
considering~$\binom{n}{N}$ parameters.

\subsection{Definition of the pair of anti-commutation matrices $\DF{}=\left(\DFM{},\DFP{}\right)$}

Because the \TRDM\ involves transitions between different wavefunctions of the Fock space based on the creation and annihilation operators, the anti-commutation relation~$\Co{i}\Ao{j} + \Ao{j}\Co{i} = \delta_{ij}$ is introduced into the scalar product $\braket{\Psi_N}{\Psi'_N} = \tau$ between two any $N$-body wavefunctions~$\ket{\Psi_N}$ and~$\ket{\Psi'_N}$ belonging to~$\mathcal{H}_N$
\begin{align}
    \mel{\Psi_N}{\Co{i}\Ao{j}  + \Ao{j}\Co{i}}{\Psi'_N} & = \tau \delta_{ij}\\
    \sum_{m}^{\binom{n}{N-1}}
        \mel{\Psi_N}{\Co{i}}{I_{N-1}^m}
        \mel{I_{N-1}^m}{\Ao{j}}{\Psi'_N}
  + \sum_{l}^{\binom{n}{N+1}}  
        \mel{\Psi_N}{\Ao{j}}{I_{N+1}^l}
        \mel{I_{N+1}^l}{\Co{i}}{\Psi'_N}
        & = \tau \delta_{ij},
\end{align}
where we introduce the two single-determinant basis sets {$\displaystyle\sum_{m}^{\binom{n}{N-1}} \ketbra{I_{N-1}^m}{I_{N-1}^m} = \mathbb{I}_{\mathcal{H}_{N-1}}$} and {$\displaystyle\sum_{l}^{\binom{n}{N+1}} \ketbra{I_{N+1}^l}{I_{N+1}^l} = \mathbb{I}_{\mathcal{H}_{N+1}}$} respectively for the Hilbert spaces~$\mathcal{H}_{N-1}$ and~$\mathcal{H}_{N+1}$.
In fact, any basis set of ~$\mathcal{H}_{N-1}$ and~$\mathcal{H}_{N+1}$ can be used but we prefer restrict to the single-determinant basis set for the simplicity of the proof.

A pair of coupled matrices $\DF{} = \left( \DFM{}, \DFP{} \right)$, which we call a pair of anti-commutation matrices (ACMP), is defined
\begin{align}
 \DFMM{}{i}{m} 
   = \DFmm{}{i}{m}
    := & \mel{I_{N-1}^m} {\Ao{i}}{\Psi_{N}} = \mel{\Psi_{N}} {\Co{i}}{I_{N-1}^m}^*\label{definition-AIM}\\
 \DFPM{}{i}{l} 
   = \DFpm{}{i}{l}
    := & \mel{I_{N+1}^l} {\Co{i}}{\Psi_{N}} = \mel{\Psi_{N}} {\Ao{i}}{I_{N+1}^l}^*,\label{definition-CIM}
\end{align}
which are linked by this relation which is the direct transposition of the anti-commutation relation~\ref{rel-ca}.
\begin{align}
%\Trans{\DF{}} \DF{\prime} = \Trans{\DFM{}} \DFM{\prime} + \Trans{\Trans{\DFP{}} \DFP{\prime}} & = \tau \mathbb{I}_n\\
\sum_m^{\binom{n}{N-1}} \DFmm{}{\alert{i}}{m}^{\alert{*}} \DFmm{\textcolor{blue}{\prime}}{\textcolor{blue}{j}}{m} + \sum_l^{\binom{n}{N+1}} \DFpm{}{\textcolor{blue}{j}}{l}^{\alert{*}} \DFpm{\textcolor{blue}{\prime}}{\alert{i}}{l}
= \pscal{\DFm{}{\alert{i}}}{\DFm{\textcolor{blue}{\prime}}{\textcolor{blue}{j}}} + \pscal{\DFp{}{\textcolor{blue}{j}}}{\DFp{\textcolor{blue}{\prime}}{\alert{i}}} 
    & = \tau \delta_{\alert{i}\textcolor{blue}{j}},
\end{align}
where we introduce the notation~$\pscal{.}{.}$ to define the scalar product between two columns of matrices.

\begin{figure}[t]
\begin{tikzpicture}[thick,scale=1.3]
    \draw (0.0,0.25) rectangle (1.0,2.75) node[pos=.50] {$\DFMN{}{N}$};
    \draw (4.0,0.25) rectangle (5.0,2.75) node[pos=.50] {$\DFPN{}{N}$};
    \node at (0.5,3.0) {$\Ao{1}\ldots\Ao{n}$};
    \node at (4.5,3.0) {$\Co{1}\ldots\Co{n}$};
    \node at (0.5,0.0) {$\DFm{}{1}\ldots\DFm{}{n}$};
    \node at (4.5,0.0) {$\DFp{}{1}\ldots\DFp{}{n}$};
    \node[left] at (0.0,-0.3) {{\tiny $M=\binom{n}{N-1}$}};
    \node[left] at (4.0,-0.3) {{\tiny $L=\binom{n}{N+1}$}};
    \node[left]  at (0.0,1.5) {$\begin{array}{c}
         \ket{I_{N-1}^1}  \\
         \vdots\\
         \ket{I_{N-1}^m}  \\
         \vdots\\
         \ket{I_{N-1}^M}
    \end{array}$};
    \node[left]  at (4.0,1.5) {$\begin{array}{c}
         \ket{I_{N+1}^1}  \\
         \vdots\\
         \ket{I_{N+1}^l}  \\
         \vdots\\
         \ket{I_{N+1}^L}
    \end{array}$};
\end{tikzpicture}
\caption{Representation of the pair~$\DF{}=\left(\DFM{},\DFP{}\right)$ of anti-commutation matrices where $\DFM{} \in \mathcal{A}_N=\mathbb{C}^{\binom{n}{N-1}}\times\mathbb{C}^n$, $\DFP{} \in \mathcal{C}_N=\mathbb{C}^{\binom{n}{N+1}}\times\mathbb{C}^n$ and so $\DF{} \in \BigSet$.
The $j$th column~$\DFm{}{j}$ of $\DFM{}$ is a vector of size~$\binom{n}{N-1}$ representing $\Ao{j}\ket{\Psi_N}$ and the $j$th column~$\DFp{}{j}$ of size~$\binom{n}{N+1}$ representing~$\Co{j}\ket{\Psi_N}$.}
\label{Fig-AC}
\end{figure}
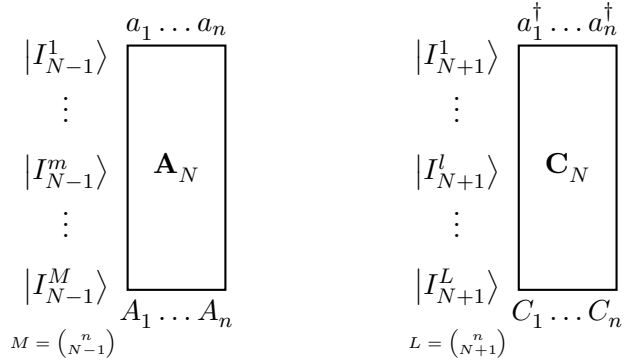

So for each wavefunction, we define two rectangular large matrices~$\DFM{}$ and $\DFP{}$ with $n$ columns and respectively $\binom{n}{N-1}$ lines and $\binom{n}{N+1}$ lines as we sketch in the figure~\ref{Fig-AC}.
The $j$th column of $\DFM{}$ represents $\Ao{j}\ket{\Psi_N}$, and the $j$th column of $\DFP{}$ $\Co{j}\ket{\Psi_N}$.
The matrix $\DFM{}$ is an element of the set $\mathcal{A}_N=\mathbb{C}^{\binom{n}{N-1}}\times\mathbb{C}^n$, the matrix $\DFP{}$ an element of the set $\mathcal{C}_N=\mathbb{C}^{\binom{n}{N+1}}\times\mathbb{C}^n$ and so $\DF{}$ belongs to the set $\BigSet$.

In order to keep the right number~$N$ of electrons, the occupation number operator~$\hat{N}=\sum_j\Co{j}\Ao{j}$ is also introduced into the scalar product given another equality 
\begin{align}
 \mel{\Psi_N}{\sum_j^n \Co{j}\Ao{j}}{\Psi_N'} = \Trace{\Trans{\DFM{}} \DFM{\prime}} = \sum_j^n \pscal{\DFm{}{j}}{\DFm{\prime}{j}} & = N\tau.
\label{relations-N-ACMP}
\end{align}

So, for any $\DF{}$ and $\DF{\prime}$ coming from $\ket{\Psi_N}$ and $\ket{\Psi'_N}$, we have, by construction from the scalar product $\braket{\Psi_N}{\Psi'_N} = \tau$, the two following equality conditions which encode the relation of anti-commutation and the number of $N$ electrons in the system
\begin{align}
\Trans{\DF{}} \DF{\prime} := \Trans{\DFM{}}\DFM{\prime} + \left( \Trans{\DFP{}}\DFP{\prime} \right)^T = \tau\mathbb{I}_n
\, , \quad
\Trace{\Trans{\DFM{}}\DFM{\prime}} & = \tau N,
\label{ACMP-conditions}
\end{align}
where we define the product as $\Trans{\DF{}} \DF{\prime}$ between two ACMPs which must be always equal to the identity matrix  $\mathbb{I}_n$ times a complex scalar.
We will show that these only two equality relations, we call ACMP conditions, are necessary and sufficient conditions to preserve the $N$-representability.
The ACMP conditions have to be checked for each ACMP, but also for any couple of ACMP which impose a very large number of strict equality conditions.
Comparing to the other \TRDM\ approach where the conditions are only inequalities, in this article, we have only equalities but with a pair of matrices~$\DF{}=\left( \DFM{}, \DFP{} \right)$.

\section{Building $N$-representable \ORDM\ and \TRDM\ from the ACMPs}

Now we can reinterpret the \ORDM\ as the scalar products between the columns of the ACMP~$\DF{0}=\left(\DFM{0},\DFP{0}\right)$ of $N$ electrons corresponding to~$\ket{\Psi_N}$ and the \TRDM\ as the scalar products between $n$ ACMPs~$\DF{i}=\left(\DFM{i},\DFP{i}\right)$ of $N-1$ electrons corresponding to~$\Ao{i}\ket{\Psi_{N}}$ coming from a $N$-body wavefunction~$\ket{\Psi_{N}}$ as
\begin{align}
    \RDMone{ii'} & = \mel {\Psi_{N}} {\Co{i}\Ao{i'}} {\Psi_{N}} = \pscal{\DFm{0}{i}}{\DFm{0}{i'}},\\
    \RDMtwo & = \mel {\Psi_{N}} {\Co{i}\Co{j}\Ao{j'}\Ao{i'}} {\Psi_{N}} = \mel {\Ao{i}\Psi_{N}} {\Co{j}\Ao{j'}} {\Ao{i'}\Psi_{N}}
            = \pscal{\DFm{i}{j}}{\DFm{i'}{j'}}.
    \label{eq:RDM-AIJ}
\end{align}
with $\DF{0}=\left(\DFM{0},\DFP{0}\right)$ is the ACMP coming from $\ket{\Psi_{N}}$ and $\DF{i}=\left(\DFM{i},\DFP{i}\right)$ is the ACMP coming from $\ket{\Ao{i}\Psi_{N}}$.
The term $\RDMtwo$ for a given $i$ can be interpreted as the \ORDM\ of the wavefunction~$\ket{\Ao{i}\Psi_{N}}$.
The corresponding total energy is so given by
\begin{equation}
    E_0 = \sum_{i;i'}^n \Hone{ii'} \pscal{\DFm{0}{i}}{\DFm{0}{i'}} + \sum_{ij;i'j'}^n \Htwo \pscal{\DFm{i}{j}}{\DFm{i'}{j'}}
    \label{Energy-ACMP}
\end{equation}
We point out that the required ACMP conditions~\ref{ACMP-conditions} and the energy expression~\ref{Energy-ACMP} are only based on scalar products between a set of ACMPs.
Contrary to the wavefunction, the information about the action of an annihilation~$\Ao{i}$ or creation~$\Co{i}$ operators into the wavefunction~$\ket{\Psi_N}$ are well localized and expressed by the $i$th column $\DFm{0}{i}$ and $\DFp{0}{i}$. 
In the same way, the information about the action of $\Ao{j}\Ao{i}$ into~$\ket{\Psi_N}$ is only stored in the $j$th column $\DFm{i}{j}$.

Thanks to the symmetry $\Ao{j}\Ao{i}\ket{\Psi_N}=-\Ao{i}\Ao{j}\ket{\Psi_N}$, because
$\DFm{i}{j}$ corresponds to $\Ao{j}\Ao{i}\ket{\Psi_N}$ and
$\DFm{j}{i}$ corresponds to $\Ao{i}\Ao{j}\ket{\Psi_N}$,
we have
\begin{align}
    \DFm{i}{j} = - \DFm{j}{i}
\end{align}
%All these information are decoupled contrary to the wavefunction~$\ket{\Psi_N}$ where we need to apply the annihilation operators~$\Ao{j}\Ao{i}$ into all Slater determinants.
The matrix product~$\RDMOne_e=\Trans{\DFM{0}}\DFM{0}$ defines the one-body electron density matrix of the corresponding wavefunction~$\ket{\Psi_N}$ and the matrix product~$\RDMOne_h=\Trans{\DFP{0}}\DFP{0}$ is the one-body hole density matrix.
In the same way, the matrix products~$\RDMOne_e^i=\Trans{\DFM{i}}\DFM{i}$ and~$\RDMOne_h^i=\Trans{\DFP{i}}\DFP{i}$ are respectively the one-body electron and hole density of the wavefunction~$\Ao{i}\ket{\Psi_N}$.
All these ACMPs can see as the decompositions (\textit{i.e.} square root) of symmetric semi-definite positive density matrices. The Cholesky decomposition is a way to do that.

So, to calculate the total energy, we have to consider a ACMP~$\DF{0}$ coming from~$\ket{\Psi_N}$ and $n$ $\DF{i}$ ACMPs coming from the $n$ $\Ao{i}\ket{\Psi_N}$ with the symmetry $\DFm{j}{i}=-\DFm{i}{j}$.
We call by $\GPF{}$ this set of $\BigSetM$  with the symmetry $\DFm{j}{i}=-\DFm{i}{j}$.
By construction, these ACMP conditions are necessary \textit{i.e.} each ACMP $\DF{0}$ and $\GPF{}$ coming from a $N$-body wavefunction obey to the ACMP conditions.

\section{Sufficiency of the ACMP conditions}

Our goal is to prove that the matrix pair $\DF{}\in\BigSet$ space of $N$-representable which respects the ACMP
conditions are $N$-representable, and also that a set $\GPF{}\in\BigSetM$ which respects the ACMP conditions are also $N$-representable.

Our demonstration will based only on the convexity and we  start first with one ACMP $\DF{}$ and then with a set $\GPF{}$ of $n$ ACMP.

\subsection{The singular value decomposition}
\label{section-enlarge-ACMP}

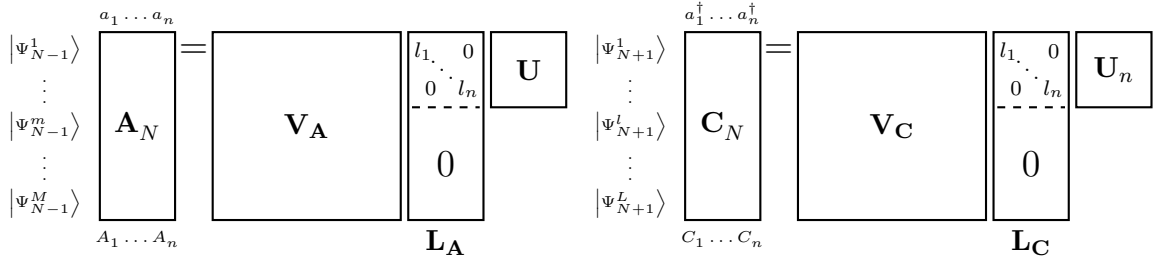
\begin{figure}[t]
    \centering
        \begin{tikzpicture}[thick]
            \draw (0.0,0.25) rectangle (1.0,2.75) node[pos=.50] {$\DFMN{}{N}$};
            \node at (0.5,3.0) {{\tiny $\Ao{1}\ldots\Ao{n}$}};
            \node at (0.5,0.0) {{\tiny $\DFm{}{1}\ldots\DFm{}{n}$}};
            %\node[left] at (0.0,0.0) {{\tiny $M=\binom{n}{N-1}$}};
            \node[left]  at (0.0,1.5) {{\tiny $\begin{array}{c}
                 \ket{\Psi_{N-1}^1}  \\
                 \vdots\\
                 \ket{\Psi_{N-1}^m}  \\
                 \vdots\\
                 \ket{\Psi_{N-1}^M}
            \end{array}$}};
            \node[right] at (0.91,2.5) {{\large $=$}};
            \draw (1.5,0.25) rectangle (4.0,2.75) node[pos=.50] {$\mathbf{V_A}$};
            \draw (4.1,0.25) rectangle (5.1,2.75);
            \draw (5.2,1.75) rectangle (6.2,2.75) node[pos=.50] {$\mathbf{U}$};
            \node at (4.3,2.5) {{\scriptsize $l_1$}};
            \node at (4.55,2.35) {{\scriptsize $\ddots$}};
            \node at (4.9,2.0) {{\scriptsize $l_n$}};
            \node at (4.9,2.5) {{\scriptsize $0$}};
            \node at (4.4,2.0) {{\scriptsize $0$}};
            \node at (4.6,1.0) {{\large $0$}};
            \node at (4.6,-0.05) {$\mathbf{L_A}$};
            %\node [above,left] at (4.1,0.50) {$\binom{n}{N-1}$};
            \draw[dashed] (4.15,1.75) -- (5.05,1.75);
        \end{tikzpicture}
        \begin{tikzpicture}[thick]
            \draw (0.0,0.25) rectangle (1.0,2.75) node[pos=.50] {$\DFPN{}{N}$};
            \node at (0.5,3.0) {{\tiny $\Co{1}\ldots\Co{n}$}};
            \node at (0.5,0.0) {{\tiny $\DFp{}{1}\ldots\DFp{}{n}$}};
            %\node[left] at (0.0,0.0) {{\tiny $M=\binom{n}{N-1}$}};
            \node[left]  at (0.0,1.5) {{\tiny $\begin{array}{c}
                 \ket{\Psi_{N+1}^1}  \\
                 \vdots\\
                 \ket{\Psi_{N+1}^l}  \\
                 \vdots\\
                 \ket{\Psi_{N+1}^L}
            \end{array}$}};
            \node[right] at (0.91,2.5) {{\large $=$}};
            \draw (1.5,0.25) rectangle (4.0,2.75) node[pos=.50] {$\mathbf{V_C}$};
            \draw (4.1,0.25) rectangle (5.1,2.75);
            \draw (5.2,1.75) rectangle (6.2,2.75) node[pos=.50] {$\mathbf{U}_n$};
            \node at (4.3,2.5) {{\scriptsize $l_1$}};
            \node at (4.55,2.35) {{\scriptsize $\ddots$}};
            \node at (4.9,2.0) {{\scriptsize $l_n$}};
            \node at (4.9,2.5) {{\scriptsize $0$}};
            \node at (4.4,2.0) {{\scriptsize $0$}};
            \node at (4.6,1.0) {{\large $0$}};
            \node at (4.6,-0.05) {$\mathbf{L_C}$};
            %\node [above,left] at (4.1,0.50) {$\binom{n}{N+1}$};
            \draw[dashed] (4.15,1.75) -- (5.05,1.75);
        \end{tikzpicture}
    \caption{Singular value decompositions of $\DFM{}$ and $\DFP{}$ where $L_A$ is a matrix of size $\binom{n}{N-1}\times n$ and $L_C$ a matrix of dimension $\binom{n}{N+1}\times n$ with $n$ diagonal non-zero terms and the other coefficients are zero, $U$ is a~$n\times n$ unitary matrix, and finally $V_A$ and $V_C$ are unitary matrices respectively of size $\binom{n}{N-1}\times\binom{n}{N-1}$ and $\binom{n}{N+1}\times\binom{n}{N+1}$.}
    \label{Fig-SVD}
\end{figure}

An ACMP~$\DF{}$ only selects $n$ vectors in the space~$\mathbb{C}^{\binom{n}{N-1}}$ and $n$ vectors in the space $\mathbb{C}^{\binom{n}{N+1}}$ \textit{i.e.} the matrices $\DFM{}$ and $\DFP{}$ are of rank~$n$.
That means that, thanks to the singular value decomposition (SVD, see figure~\ref{Fig-SVD}), an ACMP~$\DF{}$ can be decomposed into a pair $\left( V_A L_A U, V_C L_C U \right)$
where $L_A$ is a matrix of size $\binom{n}{N-1}\times n$, $L_C$ a matrix of dimension $\binom{n}{N+1}\times n$ with $n$ diagonal non-zero terms and the other coefficients are null,
\begin{align}
    \Trans{L_A} L_A + \Trans{L_C} L_C = \mathbb{I}_n, \quad
    \Trace{\Trans{L_A} L_A}=N, \quad
    \Trace{\Trans{L_C} L_C} = n-N,
    \label{L-conditions}
\end{align}
$U$ is a~$n\times n$ unitary matrix which gives the natural orbitals \textit{i.e.} the eigen-excitations of the system.
Finally $V_A$ and $V_C$ are unitary matrices respectively of size $\binom{n}{N-1}\times\binom{n}{N-1}$ and $\binom{n}{N+1}\times\binom{n}{N+1}$.
$U$ is identical for $\DFM{}$ and $\DFP{}$ due to the relation~$\Trans{\DFM{}}\DFM{}+\Trans{\DFP{}}\DFP{}=\mathbb{I}_n$.
There is no condition over the unitary matrix~$U$ and also over the unitary matrix~$V$.
If $L_A$ and $L_C$ only check the ACMP conditions~\ref{L-conditions} \textit{i.e.} $\left|(L_A)_{ii}\right| = 1$ and $\sum_i^n \left|(L_A)_{ii}\right|^2 = N$, then we have the Coleman's conditions~\cite{Coleman.1963} where the occupation number $0 \leq n_i \leq 1$ and $\sum_i^n n_i = N$ which are valid for mixed states. 
At the present stage, because only pure states \textit{i.e.} coming from a $N$-body wavefunction are considered, the Klyachko's conditions~\cite{Altunbulak-Klyachko.2008,Theophilou.2015} on $L_A$ have to be added as additional constraints which are inequalities about the occupation numbers.
By symmetry considering $n-N$ holes, $L_C$ has also additional Klyachko's conditions.

We can multiply all $\DF{}$ coming from a $N$-body wavefunction by any couple of unitary matrices $V^0=(V^0_A,V^0_C)$ which gives a set of $N$-representable ACMP where all ACMP conditions~\ref{ACMP-conditions} and the Klyachko conditions are preserved and so the pure $N$-representability.
The singular decomposition  means that any matrix pair~$\PF{} = \left( V'_A L_A U, V'_C L_C U \right)\in \BigSet$ respecting the ACMP conditions~\ref{ACMP-conditions} and Klyachko's conditions corresponds to a given wavefunction~$\ket{\Psi_N}$.

\subsection{Quotient space and compacting the ACMPs}
%\subsection{Compacting the ACMPs $\DF{}$}
\label{section-compact-ACMP}

The invariance by a unitary matrix for the ACMP $\DF{}$ gives a relation of equivalence and a way to consider a quotient space reducing the size of the ACMP matrices.
We can define a relation between two ACMPs $\DF{}$ and $\DF{\prime}\in\BigSet$, if there exists a pair of unitary matrices
$\mathbf{V}=\left(\mathbf{V_A},\mathbf{V_C}\right)\in U\left[\binom{n}{N-1}\right]\times U\left[\binom{n}{N+1}\right]$ where
\begin{align}
    \DFM{\prime} = \mathbf{V_A} \DFM{}
    \quad\text{and}\quad 
    \DFP{\prime} = \mathbf{V_C} \DFP{}
\end{align}
This relation is reflexive, symmetric and transitive given a relation of equivalence.
A quotient space of $\BigSet$ as $\SetQ=\BigSetQ$ defines a unique object for each electron \ORDM\ with a compact representation which we call by extension an ACMP.

In a same way, a \TRDM\ can be associated to a subset of $n$ ACMPs of $\SetMQ=\BigSetMQ$ which respect the ACMP conditions between each other.
A way to represent these ACMPs is to consider triangular matrices for the $\binom{n}{2}$~vectors $\DFm{i}{j}$ and the
$n^2$~vectors $\DFp{i}{j}$, which is the key to compact the ACMPs.

$\SetQ$ and $\SetMQ$ are the compact way to define ACMPs.
A compact representation of these quotient spaces gives a compact representation of the ACMPs.
The most compact representation is to use triangular matrices as we did with the Cholesky decomposition.

It is possible to use other representations as $\DF{} = \left( L_A U, L_C U \right)$
with all matrices of dimension~$n\times n$, $U$ a unitary matrix and $L_A$ and $L_C$ diagonal matrices but this is the
most compact one. 
More generally, if we have a matrix pair $\DF{} = \left( \DFM{},\DFP{} \right)$ of dimensions $m\times n$ for $\DFM{}$ and and $l\times n$ for $\DFP{}$ with respecting the ACMP conditions~\ref{ACMP-conditions}, they can always ``enlarged'' in a higher dimensional space as $\BigSet$ if $m \leq \binom{n}{N-1}$ and $l \leq \binom{n}{N+1}$.

To resume, the quotient space gives a compact representation of the ACMPs.

%\subsection{The ACMP conditions are sufficient in $\BigSet$ for the $N$-representability}
\subsection{Pure and mixed states}

\begin{figure}
    \centering
    \begin{tikzpicture}[fill=gray,fill opacity=0.7,draw,scale=3,rounded corners=0.5pt]
    	\filldraw (-0.5,-0.5,-0.5) -- ++(1,0,0) -- ++(0,1,0) -- ++(-1, 0, 0) -- cycle;
    	\filldraw (-0.5,-0.5,-0.5) -- ++(1,0,0) -- ++(0,0,1) -- ++(-1, 0, 0) -- cycle;
    	\filldraw (-0.5,-0.5,-0.5) -- ++(0,1,0) -- ++(0,0,1) -- ++(0, -1, 0) -- cycle;
    	\filldraw[fill=gray!20] (0.5,0.5,0.5) -- ++(-1,0,0) -- ++(0,-1,0) -- ++(1, 0, 0) -- cycle;
    	\filldraw[fill=gray!50!black!50] (0.5,0.5,0.5) -- ++(-1,0,0) -- ++(0,0,-1) -- ++(1, 0, 0) -- cycle;
    	\filldraw[fill=gray!20!black!80] (0.5,0.5,0.5) -- ++(0,-1,0) -- ++(0,0,-1) -- ++(0, 1, 0) -- cycle;
        \node[above right,black] at (-0.5,-0.5,-0.5) {$(0,0,0)$};
        \node[left,black]        at (-0.5,-0.5, 0.5) {$(1,0,0)$};
        \node[right,black]       at ( 0.5,-0.5,-0.5) {$(0,1,0)$};
        \node[left,black]        at (-0.5, 0.5,-0.5) {$(0,0,1)$};
        \node[left,black]        at (-0.5, 0.5, 0.5) {$(1,0,1)$};
        \node[right,black]       at ( 0.5, 0.5,-0.5) {$(0,1,1)$};
        \node[right,black]       at ( 0.5,-0.5, 0.5) {$(1,1,0)$};
        \node[above left,black]  at ( 0.5, 0.5, 0.5) {$(1,1,1)$};
        \node[black] at (2.2,0) {
            \begin{scriptsize}
                \begin{tabular}{c|c}
                    Electron \ORDM & Hole \ORDM \\ \hline
                     $(0,0,0)$ & $(1,1,1)$ \\
                     $(1,0,0)$ & $(0,1,1)$ \\
                     $(0,1,0)$ & $(1,0,1)$ \\
                     $(0,0,1)$ & $(1,1,0)$ \\
                     $(1,1,0)$ & $(0,0,1)$ \\
                     $(0,1,1)$ & $(1,0,0)$ \\
                     $(1,0,1)$ & $(0,1,0)$ \\
                     $(1,1,1)$ & $(0,0,0)$
                \end{tabular}
            \end{scriptsize}};
    \end{tikzpicture}
    \caption{
    Manifold (hypercube) of the diagonal part of the $N$-representable electron \ORDM\ for a system with $3$ states. 
    The eight vertices corresponds to the \ORDM\ of the single-determinant wavefunctions. 
    The ACMP conditions gives a relation between the vertices of electron \ORDM\ and hole \ORDM\ indicating on the right panel which correspond to the Coleman's conditions for the diagonal part.}
    \label{Fig-Hypercube}
\end{figure}
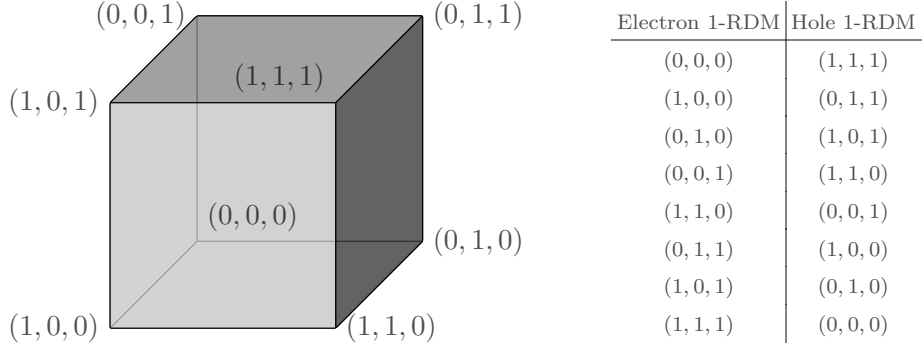

Klyachko's conditions can be removed considering only the Coleman's conditions which gives $N$-representable \ORDM\ \textit{i.e.} coming from a mixed of pure states.
%included in the ACMP conditions if we consider $\DF{}\in\SetQ$ 
The mixed-states with the associated \ORDM\ $\RDMOneMixed$ are defined as the convex set of the \ORDM\ (see the figure~\ref{Fig-Hypercube}) of the pure-state $N$-body wavefunctions. 
For $m$ wavefunctions associated with $m$ positive weights $\left(\ket{\Psi_l},\alpha_l\right)_{k=1,m}$ ($\alpha_l>0$ and $\sum_l\alpha_l=1$), the mixed-state \ORDM~$\RDMOneMixed$ are defined by
\begin{align}
    \RDMOneMixed = \sum_l^m \alpha_l \RDMone{l} = \sum_l^m \alpha_l \Trans{\DFM{l}}\DFM{l}
\end{align}
where $\DF{l}=\left(\DFM{l},\DFP{l}\right)\in\SetQ$ are the ACMP of the wavefunction~$\ket{\Psi_l}$.
Because the positivity is preserved by convexity, this electron \ORDM\ $\RDMOneMixed$ is semi-definite positive and also the corresponding hole \ORDM. 
Aa mixed-state ACMP $\DFMixed\in\SetQ$ can be defined as
\begin{align}
    \Trans{\DFMMixed}\DFMMixed =  \RDMOneMixed = \sum_l^m \alpha_l \Trans{\DFM{l}}\DFM{l} 
    \quad\text{and}\quad 
    \Trans{\DFPMixed}\DFPMixed = \sum_l^m \alpha_l \Trans{\DFP{l}}\DFP{l}
\end{align}
using a Cholesky decomposition for instance.

%Then multiplying on the left, respectively, by any unitary matrix~$V_A$ of $U\left[\binom{n}{N-1}\right]$ the unitary group of $\binom{n}{N-1}$ dimension and by any unitary matrix~$V_C$ of $U\left[\binom{n}{N+1}\right]$, we build $\DFMMixed$ and a $\DFPMixed$
Moreover, the ACMP conditions are preserved by linearity and so by convexity
\begin{align}
    \Trans{\DFMMixed}\DFMMixed + \Trans{\DFPMixed}\DFPMixed 
    & = \sum_l^m \alpha_l \left( \Trans{\DFM{l}}\DFM{l} + \Trans{\DFP{l}}\DFP{l} \right)
      = \sum_l^m \alpha_l \mathbb{I}_n = \mathbb{I}_n\nonumber\\
    \Trace{\Trans{\DFMMixed}\DFMMixed}
    & = \Trace{\RDMOneMixed} = \sum_l^m \alpha_l \Trace{\Trans{\DFM{l}\DFM{l}}} = \sum_l^m \alpha_l N = N.
    \label{ACMP-convexity}
\end{align}
In that case, the corresponding $\DFMixed=\left(\DFMMixed,\DFPMixed\right)$ can not be in correspondence with a given wavefunction but the \ORDM\
$\Trans{\DFMMixed}\DFMMixed$ and the $\Trans{\DFPMixed}\DFPMixed$ are in the convex hull of the pure $N$-representable~$\DF{}$~\cite{Coleman.1963}.
Thanks to the singular value decomposition, because any $\DF{}\in\SetQ$ respecting the ACMP conditions, checks the Coleman's conditions, the corresponding \ORDM\ is in the hypercube.

On the contrary, if a weight~$\alpha_l$ is negative, the positivity of the \ORDM\ $\RDMOneMixed$ is not
preserved and a corresponding $\DFMixed$ can not be defined by a Cholesky decomposition for instance.
So the ACMP conditions are sufficient. Any ACMP $\DF{}\in\SetQ$ checking only the ACMP conditions~\ref{ACMP-conditions} are $N$-representable.
Only the Coleman's conditions have to be considered which is easier to handle than Klyachko's conditions.

\subsection{Sufficiency of the ACMP conditions in $\BigSet$ for $N$-representable \TRDM}

The final step is to prove the $N$-representability of a set~$\GPF{}\in\SetMQ$ of $n$ $\DF{i}$ with property $\DFm{i}{j} = -\DFm{j}{i}$.
In the same way as for the \ORDM, we consider the convex hull of the pure-state \TRDM.
For $m$ sets associated with $m$ positive weights~$\left(\GPF{l},\alpha_l\right)_{l=1,m}$ ($\alpha_l>0$ and
$\sum_l\alpha_l=1$) coming from $m$ wavefunctions~$\ket{\Psi_l}$, we can define the corresponding mixed-states 2-RDM~$\RDMtwoMixed$ as
\begin{align}
    \RDMtwoMixed 
    & = \sum_l^m \alpha_l \left[\mathbf{{}^2\Gamma_l}\right]^{ii'}_{jj'} 
      = \sum_l^m \alpha_l \pscal{\DFm{i,l}{j}}{\DFm{i',l}{j'}}.
\end{align}
The ACMP conditions are also preserved by convexity as shown by the relation~\ref{ACMP-convexity}.
We can also define a corresponding mixed-state $\GPFMixed$.
For that we consider the associated overlap square matrix~$\mathbf{S_A}$ of size $\binom{n}{2}$ defined by the scalar
products~$\pscal{\DFm{i}{j}}{\DFm{i'}{j'}}$ between the $\binom{n}{2}$ vectors $\DFm{i}{j}$ and the overlap square matrix~$\mathbf{S_C}$ of size $n^2$ defined by the scalar products~$\pscal{\DFp{i}{j}}{\DFp{i'}{j'}}$.
By construction the overlap matrices $\mathbf{S_A}$ and $\mathbf{S_C}$ are also semi-definite positive.

Because the convexity preserves the positivity of the overlap matrices
$\mathbf{S^A}$ and $\mathbf{S^C}$, we can build $\binom{n}{2}$ vectors $\DFm{i,\text{m}}{j}$ and $n^2$ vectors
$\DFp{i,\text{m}}{j}$ where the corresponding overlap matrices are the barycenter of $m$ set~$\GPF{l}$ as
\begin{align}
    \pscal{\DFm{i,\text{m}}{j}}{\DFm{i',\text{m}}{j'}}
        & = \left[ \mathbf{S_A^m} \right]_{(i'j'),(ij)}
            = \sum_l^m \alpha_l \left[ \mathbf{S_A^l} \right]_{(i'j'),(ij)}
            = \sum_l^m \alpha_l \pscal{\DFm{i,l}{j}}{\DFm{i',l}{j'}}\\
    \pscal{\DFp{i,\text{m}}{j}}{\DFp{i',\text{m}}{j'}}
        & = \left[ \mathbf{S_C^m} \right]_{(i'j'),(ij)}
            = \sum_l^m \alpha_l \left[ \mathbf{S_C^l} \right]_{(i'j'),(ij)}
            = \sum_l^m \alpha_l \pscal{\DFp{i,l}{j}}{\DFp{i',l}{j'}} 
\end{align}
The Cholesky decomposition can be used to define both matrices, and then we have a set $\GPFMixed\in\SetMQ$.

For the sufficiency, we prove the reciprocal by contradiction.
If we take a set~$\GPF{m}$ satisfying the ACMP conditions which is not $N-$representable, the resulting \TRDM\ $\RDMTwoMixed$ is the linear combination of the \TRDM\ of pure states with at least one associated negative weight $\alpha_l$.
%The resulting \TRDM\ $\RDMTwoMixed$ can be semi-definite positive because we have not established the preservation of the positivity for the \TRDM.

In that case, according to the previous section, the corresponding {\ORDM}s $\left[\mathbf{{}^1\Gamma^i_{\text{m}}}\right]_{jj'} = \sum \alpha_l
\pscal{\DFm{i,l}{j}}{\DFm{i,l}{j'}}$ are not semi-definite positive because a weight $\alpha_l$ is negative so we can not define a corresponding $\DFm{i}{\text{m}}$ which is in contradiction with our starting assumption.

This means that the ACMP conditions are also sufficient to have a $N$-representable set~$\GPF{}$.
This concludes the statement that the ACMP conditions are necessary and sufficient to have a $N$-representable set $\GPF{}\in\SetMQ$.

In conclusion, any ACMPs with the properties~\ref{ACMP-conditions}, even of small dimensions, are $N$-representable. We only have all information for the application of one annihilation or creation operator and we totally lost information about applications with two or more annihilation and creation operators. 
In the case of the \TRDM, we consider a set $\GPF{}$ of $n$ matrix pairs~$\DF{i}$ with the symmetry $\DFm{i}{j}=-\DFm{j}{i}$. 
This anti-symmetry is preserved when we multiply all matrix pairs by the same unitary matrix pair or when we enlarge the matrices.
This means that we need to consider $\binom{n}{2}$ columns~$\DFm{i}{j}$ and $n^2$ columns $\DFp{i}{j}$ given a size for the $\DFm{i}{j}$ as $\binom{n}{2}$ and $n^2$ for the size of $\DFp{i}{j}$.

\section{Lagrangian and numerical aspects}

To minimize the total energy, we consider all sets $\GPF{}$ of~$n$ ACMPs $\DF{i}$ of $\SetMQ$ normalized by a $\DF{0}\in\SetQ$ with
\begin{align}
    \pscal{\DF{i}}{\DF{i'}} & = \RDMone{ii'} = \pscal{\DFm{0}{i}}{\DFm{0}{i'}},\quad\Trace{\Trans{\DFM{i}}\DFM{i'}} = (N-1)\RDMone{ii'}\quad\text{and}\quad\DFm{i}{j} = -\DFm{j}{i}.
\end{align}
The minimal sizes of the matrices are $min\left[\binom{n}{2},\binom{n}{N-2}\right]\times n$ for $\DFM{i}$ and $min\left[n^2,\binom{n}{N}\right]\times n$ for $\DFP{i}$ which are really small compared to the dimension~$\binom{n}{N}$ of~$\mathcal{H}_N$.
For~$\DF{0}$, $\DFM{0}$ has a size of $min\left[n,\binom{n}{N-1}\right]\times n$ and $\DFP{0}$ a size of $min\left[n,\binom{n}{N+1}\right]\times n$. Matrices with larger dimensions do not preserve the $N$-representability.
%$\DF{0}$ represents "a volume" of a set of~$n$ ACMPs of $\mathcal{D}_{N-1}$.
%
% \begin{figure}
%  \includegraphics[width=0.5\textwidth]{Figure-2RDM.pdf}
%  \caption{Sketch of the subspace of $n$ ACMP.}
% \end{figure}
%
%\section{Lagrange multipliers and Euler-Lagrange equations}
%

In order to impose the equality constraints, we use the Lagrange multipliers approach to minimize the ground state energy with the constraints to have a set of $n$ ACMPs for $N-1$ electrons.

\begin{align}
    & L \left[ \DF{0}, \left\{ \DF{i} \right\}_{i=1,n},
    \mathbf{\lambda}, \mathbf{\Lambda}, \mu_0, \mathbf{\mu} \right]
    = \sum_{ii'}^n \Hone{ii'} \pscal{\DFm{0}{i}}{\DFm{0}{i'}}
    + \sum_{ii';jj'}^n \Htwo \pscal{\DFm{i}{j}}{\DFm{i'}{j'}}\nonumber\\
    & - \underbrace{\sum_{ii'}^n \lambda_{ii'} \left[ \pscal{\DFm{0}{i}}{\DFm{0}{i'}} + \pscal{\DFp{0}{i'}}{\DFp{0}{i}} - \delta_{ii'}\right]}_{\text{To have}\,  \Trans{\DF{0}} \DF{0} = \,\mathbb{I}_n}
    - \underbrace{\sum_{ii';jj'}^n \Lambda^{ii'}_{jj'} \left[  \pscal{\DFm{i}{j}}{\DFm{i'}{j'}} + \pscal{\DFp{i}{j'}}{\DFp{i'}{j}} - \delta_{jj'}\pscal{\DFm{0}{i}}{\DFm{0}{i'}}\right]}_{\text{To have}\, \Trans{\DF{i}} \DF{i'} = \,\RDMone{ii'}\mathbb{I}_n}\nonumber\\
    & - \underbrace{\mu_0 \left[ \sum_j^n \pscal{\DFm{0}{j}}{\DFm{0}{j}} - N \right]}_{\text{To have}\, N\,\text{electrons}}
    - \underbrace{\sum_{ii'}^n \mu_{ii'} \left[ \sum_j \pscal{\DFm{i}{j}}{\DFm{i'}{j}} - (N-1)\pscal{\DFm{0}{i}}{\DFm{0}{i'}}\right]}_{\text{To have}\, \RDMone{ii'}(N-1)\,\text{electrons for }\,Tr(\Trans{\DFM{i}}\DFM{i'})}.
\end{align}
The calculation of the two-body energy term and some conditions needs $n^4$ scalar products of size $n^2$ which gives a method of $\mathcal{O}\left(n^6\right)$~complexity.

\begin{table}[ht]
     \centering
     \begin{tabular}{c|c|c|c|c|c|c|c|c}
     n   & N   & Energy        & \ORDM         & \TRDM         & $\DF{0}$      & $\DFM{0}$     & $\DF{i}$      & $\DFM{i}$ \\ \hline
     $3$ & $2$ & $1.0\times10^{-11}$ & $2.9\times 10^{-11}$ & $3.0\times 10^{-11}$ & $2.1\times 10^{-12}$ & $3.4\times 10^{-13}$ & $1.3\times 10^{-12}$ & $6\times 10^{-13}$\\
     $5$ & $2$ & $5.6\times10^{-4}$ & $9.0\times 10^{-4}$  & $4.8\times 10^{-3}$  & $1.3\times 10^{-6}$  & $7.4\times 10^{-7}$  & $7.3\times 10^{-6}$  & $7.9\times 10^{-6}$\\
     $7$ & $4$ & $9.0\times10^{-4}$ & $9.5\times 10^{-4}$  & $4.4\times 10^{-3}$  & $4.6\times 10^{-6}$  & $1.6\times 10^{-6}$  & $2.3\times 10^{-5}$  & $3.2\times 10^{-5}$\\
     $8$ & $4$ & $1.7\times10^{-2}$ & $4.8\times 10^{-2}$  & $4.3\times 10^{-2}$  & $1.1\times 10^{-5}$  & $3.4\times 10^{-6}$  & $3.1\times 10^{-5}$  & $5.7\times 10^{-5}$\\
     $9$ & $2$ & $3.6\times10^{-3}$ & $1.7\times 10^{-3}$  & $5.7\times 10^{-3}$  & $3.6\times 10^{-6}$  & $3.2\times 10^{-6}$ & $1.6\times 10^{-5}$  & $2.3\times 10^{-5}$\\
     $9$ & $3$ & $4.0\times10^{-3}$ & $2.6\times 10^{-3}$  & $6.6\times 10^{-3}$  & $8.8\times 10^{-6}$  & $3.4\times 10^{-6}$  & $4.9\times 10^{-5}$  & $8.6\times 10^{-5}$\\
     $9$ & $5$ & $2.9\times10^{-3}$ & $2.6\times 10^{-3}$  & $6.6\times 10^{-3}$  & $8.8\times 10^{-6}$  & $3.4\times 10^{-6}$  & $4.9\times 10^{-5}$  & $8.6\times 10^{-5}$\\
     $9$ & $8$ & $1.6\times10^{-10}$ & $7.1\times 10^{-10}$  & $1.2\times 10^{-9}$  & $7.2\times 10^{-10}$  & $1.8\times 10^{-10}$  & $1.9\times 10^{-10}$  & $9.6\times 10^{-11}$\\
     $11$ & $3$ & $6.3\times10^{-3}$ & $3.1\times 10^{-3}$  & $6.1\times 10^{-3}$  & $1.2\times 10^{-5}$  & $4.5\times 10^{-6}$  & $5.3\times 10^{-5}$  & $1.0\times 10^{-4}$\\
     $12$ & $5$ & $4.1\times10^{-3}$ & $5.0\times 10^{-3}$  & $7.1\times 10^{-3}$  & $4.7\times 10^{-5}$  & $1.2\times 10^{-5}$  & $1.9\times 10^{-4}$  & $4.3\times 10^{-4}$\\
     $12$ & $9$ & $3.7\times10^{-2}$ & $4.1\times 10^{-2}$  & $4.8\times 10^{-2}$  & $8.7\times 10^{-5}$  & $7.5\times 10^{-5}$  & $1.5\times 10^{-4}$  & $4.1\times 10^{-4}$\\
     $15$ & $6$ & $2.3\times10^{-2}$ & $2.9\times 10^{-2}$  & $2.4\times 10^{-2}$  & $1.4\times 10^{-4}$  & $6.2\times 10^{-5}$  & $3.1\times 10^{-4}$  & $1.1\times 10^{-3}$\\
     $15$ & $10$ & $3.4\times10^{-2}$ & $1.3\times 10^{-2}$  & $1.4\times 10^{-2}$  & $3.1\times 10^{-4}$  & $8.8\times 10^{-5}$  & $3.7\times 10^{-4}$  & $1.5\times 10^{-4}$
     \end{tabular}
     \caption{Absolute error for the total energy, and average errors for the \ORDM\, the \TRDM\, and the four constraints about $\DF{0}$, $\DFM{0}$, $\DF{i}$, and $\DFM{i}$ for different number of states and electrons. We consider $10$ different sets of parameter for each $n$ and $N$ values.}
     \label{Table-results}
\end{table}

To test our methods, we use the python package \texttt{PySCF}~\cite{PySCF2020} which can calculate the full configuration interaction ground state energy for any two-body Hamiltonian. 
We compare our method~\cite{myrepository} programming using \texttt{numpy}~\cite{harris2020array} for the linear algebra operations and \texttt{scipy}~\cite{2020SciPy-NMeth} to find the roots of the Lagrangian gradients with the Newton-Krylov algorithm.
We have also used Algencan~\cite{Algencan} by means of julia~\cite{bezanson2017julia} language and the JuMP module~\cite{DunningHuchetteLubin2017}.
The aim is to give a numerical proof of this method which is the final test in addition to the mathematical demonstration explained in this article.
The convergence is poor but for any systems we tested up to $15$ states and any number of electrons without spin.
A good initial guess is really important and the convergence can stop with very poor results but always with a higher energy than the full CI one.

We have improved the convergence using a symmetrized version of the Lagrangian as 

\begin{align}
    & L \left[ \DF{0}, \left\{ \DF{i} \right\}_{i=1,n},
    \mathbf{\lambda}, \mathbf{\Lambda}, \mu_0, \mathbf{\mu} \right]\nonumber\\
    & = \sum_{ii'}^n \HTone{ii'} \left( \pscal{\DFm{0}{i}}{\DFm{0}{i'}}  - \pscal{\DFp{0}{i}}{\DFp{0}{i'}} \right)
    + \sum_{ii';jj'}^n \HTtwo  \left( \pscal{\DFm{i}{j}}{\DFm{i'}{j'}} - \pscal{\DFp{i}{j'}}{\DFp{i'}{j}} \right) \nonumber\\
    & -\sum_{ii'}^n \lambda_{ii'} \left[ \pscal{\DFm{0}{i}}{\DFm{0}{i'}} + \pscal{\DFp{0}{i'}}{\DFp{0}{i}} - \delta_{ii'}\right]
    - \sum_{ii';jj'}^n \Lambda^{ii'}_{jj'} \left[  \pscal{\DFm{i}{j}}{\DFm{i'}{j'}} + \pscal{\DFp{i}{j'}}{\DFp{i'}{j}} - \delta_{jj'}\pscal{\DFm{0}{i}}{\DFm{0}{i'}}\right]\nonumber\\
    & - \mu_0 \left[ \sum_j^n \pscal{\DFm{0}{j}}{\DFm{0}{j}} - \pscal{\DFp{0}{j}{\DFp{0}{j}}} - (2N-n) \right] \nonumber\\
    & - \sum_{ii'}^n \mu_{ii'} \left[ \sum_j \pscal{\DFm{i}{j}}{\DFm{i'}{j}} - \sum_j \pscal{\DFm{i}{j}}{\DFm{i'}{j}} -
    2\left( (N-1) -n \right)\pscal{\DFm{0}{i}}{\DFm{0}{i'}}\right].
\end{align}
where $\tilde{h}_{ii'} = \frac{1}{2}\Hone{ii'} + \frac{1}{4}\sum_j H^{ii'}_{jj}$ and $\HTtwo = \frac{1}{2}\Htwo$.
If we start from the full CI solution, 
we have an accuracy of less than $10^{-6}$ over the energy and over the components of the \ORDM\ and the \TRDM.
We start from the solution of the one-body Hamiltonian where we include the one-body part of the two-body Hamiltonian as
\begin{align}
    h_{ii'}^{1\rightarrow 2} & = \frac{1}{2n} \sum_j^n H2^{ii'}_{jj}\nonumber\\
    h'_{ii'} & = h_{ii'} +  h_{ii'}^{1\rightarrow 2},  
    H'^{ii'}_{jj'} = H^{ii'}_{jj'} - \frac{1}{N-1} \left( h_{ii'}^{1\rightarrow 2}\delta_{jj'} + h_{jj'}^{1\rightarrow 2}\delta_{ii'} \right)
\end{align}
Thanks to the relation $\sum_j \pscal{\DFm{i}{j}}{\DFm{i'}{j}} = (N-1)\pscal{\DFm{0}{i}}{\DFm{0}{i'}}$, the total energy is not changed and the one-body part has a larger amplitude. 
This is also a way to include the one-body into the two-body part of the Hamiltonian.

In the table~\ref{Table-results}, we show the average errors for the total energy for $10$ different calculations using $200$ iterations of the Newton-Krylov algorithm in function of the number of states and electrons.
The initial guess is the solution of the one-body Hamiltonian part.
With $N=2$ electrons, the convergence is really fast using 10 iterations for a high accuracy.
This is also the case when $N=n-1$ which is similar to the case of $N=1$ because we have a hole-electron symmetry in this formalism.

\section{Discussions}

This new method is the first exact one with a compact representation.
By construction, the semi-definite positivity of the \TRDM\ given by $\pscal{\DFm{i}{j}}{\DFm{i'}{j'}}$ and the semi-definite positivity of the particle-hole ${}^2G$ density matrix given by $\pscal{\DFp{i}{j}}{\DFp{i'}{j'}}$ are preserved which are the $N$-representable D and G conditions.
In a next article, we will give more details about the mathematics associated to the ACMPs showing that they are
isomorph to the many-body wavefunctions
The wavefunctions have redundant information and the ACMP is a way to compact this keeping only the one-body and two-body correlations.
%Moreover we think that the wavefunctions are not physical objects because it is possible to have wavefunctions solution of more than two-body interactions.
This article does not address the numerical work about convergence, efficient parametrization of ACMPs could be developed using the group theory to determine some class of solutions.
We think that these ACMPs encode the fermionic part in an elegant way. It could be possible also to generalize this formalism to boson using in this case the commutation relation.
Excited states are also easily calculated imposing the orthogonality with the ground state ACMP as $\pscal{\DF{0}}{\DF{\text{excited}}}=0$.

The generalization to the quantum field theory using the field operator has to be done considering the action of the creation and annihilation operators. 
The use of the number of electrons can be avoided which is important in the thermodynamic limit for infinite system as in the solid state physics. 
In this case it is important to consider a finite volume and all quantities are well defined. The ACMPs can replace the use of the wavefunction for infinite systems which is problematic.
That means that the grand canonical where the number of particles is not fixed can be easily investigated.

We have also nested ACMPs: a set $\GPF{}\in\SetMQ$ of ACMPs normed by~$\DF{0}\in\SetQ$ encodes the physical two-body information related to a $N$-body wavefunction. 
The three-body density matrix can also be calculated considering then another set of $n^2$ ACMPs of $N-2$ electrons.
This means also that we have constructed, in a geometrical way, the exact density functional for any two-body Hamiltonian. 
If we consider a one-body density matrix, we determine  the corresponding ACMP $N$-electron $\DF{0}$ and then we can minimize a set of $n$ ACMPs $(N-1)$-electron $\DF{i}$ in order to have the lowest energy for the given two-body  Hamiltonian.

In conclusion, in this article, we have defined a new formalism more physical than the wavefunctions, with the advantage to consider compact objects in a polynomial time to calculate not only the ground state but also the excited states.

\section{Acknowledgments}

We would like to thank especially I.~Duchemin for the fruitful discussions and all remarks about the linear algebra, and thank also L.~Genovese, G.~Carleo, R.~Rossi, and N.~Deutsch.

\bibliography{main}

\end{document}